# Reliable Mining of Automatically Generated Test Cases from Software Requirements Specification (SRS)

Lilly Raamesh[1] and G. V. Uma[2]

[1] Research Scholar, Anna University,
Chennai 25, India

[2] Asst. Professor/CSE, Anna University,
Chennai-25, India

**Abstract**
Writing requirements is a two-way process. In this paper we use to classify Functional Requirements (FR) and Non Functional Requirements (NFR) statements from Software Requirements Specification (SRS) documents. This is systematically transformed into state charts considering all relevant information. The current paper outlines how test cases can be automatically generated from these state charts. The application of the states yields the different test cases as solutions to a planning problem. The test cases can be used for automated or manual software testing on system level. And also the paper presents a method for reduction of test suite by using mining methods thereby facilitating the mining and knowledge extraction from test cases.

## 1. Introduction

The systematic production of high-quality software, which meets its specification, is still a major problem. Although formal specification methods have been around for a long time, only a few safety-critical domains justify the enormous effort of their application. The state of the practice, which relies on testing to force the quality into the product at the end of the development process, is also unsatisfactory. The need for effective test automation adds to this problem, because the creation and maintenance of the test ware is a source of inconsistency itself and is becoming a task of comparable complexity as the construction of the code.

Data mining algorithms can be applied at different levels of abstraction and help the user discover more meaningful patterns. Data mining will create patterns from the existing database. Using well-established data mining techniques, practitioners and researchers can explore the potential of this valuable data in order to manage their project and to produce higher quality software systems that are delivered on time and within budget.

## 2. Our Approach

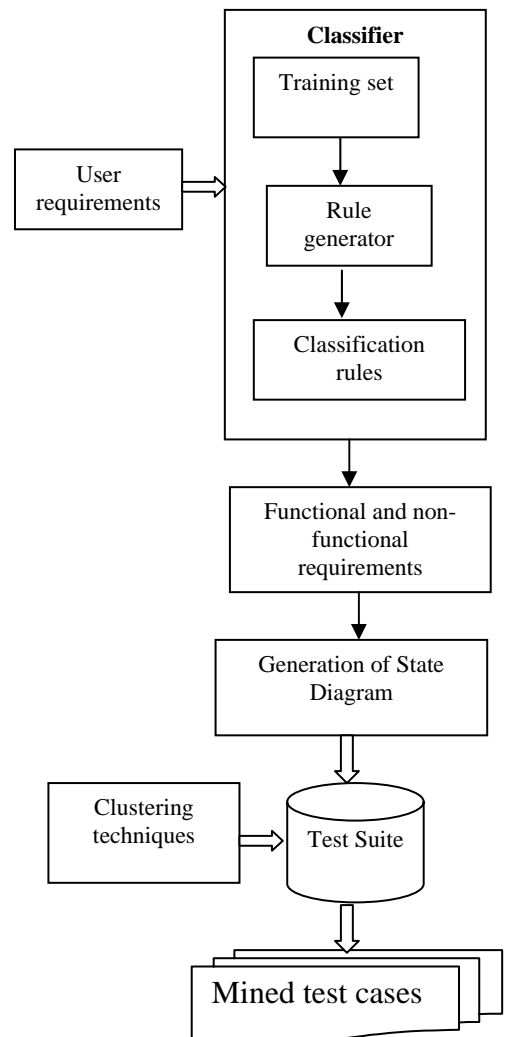

Fig 1. Automatically Generated Test Cases From Software Requirements Specification mining System

Our approach is as follows
    i. Generation of classification rules.
    ii. Generate test cases from the UML state machine.
    iii .Finally data mining techniques are applied on the generated test cases in order to further reduce the test suite size.

## 2. Generation of Classification Rules

In the current paper, we provide the Software Requirements Specification to the classifier system. For classifying we use Weka. The Weka classifier is initially trained with a training set. Later it is provided with the SRS. It classifies SRS in to functional and non functional requirements by generating a classification rules. The classification rules are applied to the SRS to get FR and NFR.
From NFR we derive the state machine. State machines specify the behaviour of a system/subsystem.

## 3. Generation of Test Cases

This section briefly describes a transformation from State diagrams in to Test cases. State machines and state diagrams have a long history in computer science. Recent versions of UML include an expressive state diagrams concept. Especially the abstraction mechanisms in the UML state machine formalism, i.e. nesting of states and stubs, allow us to map all the important elements of our use case documents to State machines.

*From State Machines to Test Cases*

Using state models to derive test cases has been common practice in the software testing world for some time .The final goal of model-based testing is to automate the test case generation from test models as much as possible. Our approach generates a set of valid test sequences, where the preconditions of all transitions are established either by previous actions or by properties of the test data. The scope of our method is the generation of test sequences supplemented by constraints on the test data, as far as these can be derived from the information presenting the state machine.

The method given in [3] can be used to create test cases from state machine diagrams which is as follows:

There are three main steps in test case generation, in the first step a predicate is selected on a transition from a UML state machine diagram. In the next step, the selected predicate is transformed into a predicate function. In the third step, test data are generated corresponding to the transformed predicate function. The generated test data are stored for use with an automatic tester. Once the test data corresponding to a particular predicate are determined, the steps are repeated by selecting the next predicate on the state machine diagram. The process is repeated until all Predicates on the state machine diagram have been considered.

*3.1. Predicate selection:*

For selecting a predicate, a traversal of the state diagram is performed using depth first (DFS) traversal or breadth first (BFS) traversal to see that every transition is considered for predicate selection. DFS traversal is used here. During traversal, conditional predicates on each of the transitions are looked. Corresponding to each conditional predicate, test data are generated.

The test data are generated for each predicate corresponding to the true or false values of the conditional predicate satisfying the prefix path condition.

*3.2. Predicate transformation:*

Let I0 consists of all variables that affect a predicate q in the path P of a state machine diagram, then two points named ON and OFF for a given border satisfying the boundary-testing criterion are created. The relational expressions of the predicates are transformed into a function F called predicate function. If the predicate q is of the form (E1 op E2), where E1 and E2 are arithmetic Expressions and op is a relational operator; then F = (E1 -E2) or (E2 - E1) depending on whichever is positive for the data I0. Next, the input data I0 is modified such that the function F decreases and finally turns negative. When F turns negative, it corresponds to the alternation of the outcome of the Predicate. Hence, as a result of the predicate transformation, the point at which the outcome of a predicate q changes, corresponds to the problem of minimization of the corresponding function F. This minimization is achieved through repeated modification of the input data values.

*3.3. Test data generation*:

The basic search Procedure we use for finding the minimum of a predicate function F is the alternating variable method. This method is based on minimizing F with respect to each input variable in turn. An initial set of inputs can be randomly generated by instantiating the data variables. Each input data variable xi is increased/ decreased in steps of Sxi, while keeping all other data variables unchanged. Here, Sxi refers to a unit step of the variable xi. The exact value of unit step can be defined conveniently. For example, unit step of 1 is normally used for integer values. Unit step can easily be defined for many other types of data such as float, double, array, and pointer and so on.

However, the method may not work when the variable assumes only a discrete set of values. Each predicate in a path can be considered to be a constraint. A path will not be traversed for some input data value, if the corresponding constraint is not satisfied. If a path P is not traversed for some data value, then we say that a constraint violation has taken place for that data value. We compute the value of F when each input datum is modified by Sxi. If the function F decreases for the modified data, and Constraint violation does not occur, then the given data variable and the appropriate direction is selected for minimising F further. Here, appropriate direction refers to whether we increase or decrease the data variable xi so that F is minimised. We start searching for a minimum with one input variable, while keeping all other input variables constant, until the solution is found (the predicate function becomes negative) or the positive minimum of the predicate function is located. In the latter case, the search continues from this minimum value with the next input variable. Two data values Iin (inside boundary) and Iout (outside boundary) are generated using the search procedure mentioned. These two points are on different sides of the boundary.

For finding these two data points, a Series of moves is made in the same direction determined by the search procedure mentioned above and the value of F is computed after each move. The size of the step is doubled after each successful move. This makes the search for the test data quick. A successful move is one where the value computed by the predicate function F is reduced. When the Minimisation function becomes negative (or zero), the required data values Iin and Iout are noted. These Points are refined further to generate a data value, which corresponds to a minimum value of the minimisation function along the last processed Direction. This refinement is done by reducing the size of the step and comparing the value of F with the previous value. Also, the distance between the data points is minimised by reducing the step size. For each Conditional predicate in the state machine diagram, we generate the test data. The generated test data are stored in a file. A test executor can use these test cases later for automatic testing.

The above said procedure produces a test suit that is of some what smaller size. But we can further reduce the size by using mining techniques.

## 4. Mining Techniques For Test Suite Reduction

Data mining is the process of extracting patterns from data. As more data are gathered, with the amount of data doubling every three years, data mining is becoming an increasingly important tool to transform these data into information. It is commonly used in a wide range of profiling practices, such as marketing, surveillance, fraud detection and scientific discovery.

While data mining can be used to uncover patterns in data samples, it is important to be aware that the use of non-representative samples of data may produce results that are not indicative of the domain. Similarly, data mining will not find patterns that may be present in the domain, if those patterns are not present in the sample being "mined". There is a tendency for insufficiently knowledgeable "consumers" of the results to attribute "magical abilities" to data mining, treating the technique as a sort of all-seeing crystal ball. Like any other tool, it only functions in conjunction with the appropriate raw material: in this case, indicative and representative data that the user must first collect. Further, the discovery of a particular pattern in a particular set of data does not necessarily mean that pattern is representative of the whole population from which that data was drawn. Hence, an important part of the process is the verification and validation of patterns on other samples of data.

The term data mining has also been used in a related but negative sense, to mean the deliberate searching for apparent but not necessarily representative patterns in large numbers of data. To avoid confusion with the other sense, the terms *data dredging* and *data snooping* are often used. Note, however, that dredging and snooping can be (and sometimes are) used as exploratory tools when developing and clarifying hypotheses.[6]

4.1. Applying data mining concepts

There are many methods available for mining different kinds of data, including association rule, characterization, classification, clustering, etc.

We can utilize any of these techniques based on

- What kind of data bases to work on
- What kind of knowledge to be mined
- What kind of techniques to be utilized

We can apply association or clustering techniques for test case mining.

### 4.1.1. ASSOCIATION

Association rules describe the association among items in the large data base. For example, one may find, from a large set of transaction data, such as association rule as if a customer buys(one brand of) milk, he/ she usually buys(another brand of) bread in the same transaction. Using these association rules, we can derive the association patterns from large databases.

### 4.1.2. DATA CLASSIFICATION

Data classification is the process which finds the common properties among a set of objects in a database and classifies them into different classes, according to a classification model.

### 4.1.3. CLUSTERING

Clustering is the process of grouping the data into classes or clusters so that object within a cluster has high similarity in comparison to another, but is dissimilar to object in other clusters. It doesn't require the class label information about the data set because it is inherently a data driven approach. It is the process of grouping or abstract object into classes of similar object.

Among all the mining techniques, clustering is the most effective technique which we are going to use for test case mining.

Clustering analysis helps constant meaningful partitioning of a large set of object based on a "divide and conquer" methodology which decomposes a large scale system into smaller components to simplify design and implementation. As a data mining task, data clustering identifies cluster or densely populated regions, according to some distance measurement, in a large, multidimensional data. Given a large set of multidimensional data points, the data space is usually not uniformly occupied by the data points. Data clustering identifies the sparse and the crowded places, and hence discovers the overall distributions patterns of the data set.

For cluster analysis to work efficiently and effectively as many literatures have presented, there are the following typical requirements of clustering in data mining.

- o Scalability:
- o Ability to deal with different types of attributes:
- o Discovery of clusters with arbitrary shape:
- o Minimal requirements for domain knowledge to determine input parameters:
- o Ability to deal with noisy data:
- o Insensitivity to the order of input records:
- o High dimensionality:

## 5. Experiments

Total 365 sentences
- • 235 annotated as "NFR"
- • 130 annotated as "FR".

| Training set no | Correctly classified sentences | In-Correctly classified sentences |
|---|---|---|
| 1 | 215 | 150 |
| 2 | 259 | 106 |
| 3 | 356 | 9 |

## 6. Conclusions

In this paper, a new approach to automatically generate test cases from SRS and mining of test cases has been discussed. Firstly a formal transformation of a detailed SRS to a UML state model, secondly the generation of test cases from the state model and lastly mining of Test cases. The introduction of agents can bring enhancement.